# Constraints and Variables Reduction for Optimal Power Flow Using Hierarchical Graph Neural Networks with Virtual Node-Splitting

Thuan Pham
*Department of Electrical and Computer Engineering*
University of Houston
Houston, TX, USA
tdpham7@cougarnet.uh.edu

Xingpeng Li
*Department of Electrical and Computer Engineering*
University of Houston
Houston, TX, USA
xli82@uh.edu

*Abstract*— Power system networks are often modeled as homogeneous graphs, which limits the ability of graph neural network (GNN) to capture individual generator features at the same nodes. By introducing the proposed virtual node-splitting strategy, generator-level attributes like costs, limits, and ramp rates can be fully captured by GNN models, improving GNN's learning capacity and prediction accuracy. Optimal power flow (OPF) problem is used for real-time grid operations. Limited timeframe motivates studies to create size-reduced OPF (ROPF) models to relieve the computational complexity. In this paper, with virtual node-splitting, a novel two-stage adaptive hierarchical GNN is developed to (i) predict critical lines that would be congested, and then (ii) predict base generators that would operate at the maximum capacity. This will substantially reduce the constraints and variables needed for OPF, creating the proposed ROPFLG model with reduced monitor lines and reduced generator-specific variables and constraints. Two ROPF models, ROPFL and ROPFG, with just reduced lines or generators respectively, are also implemented as additional benchmark models. Case studies show that the proposed ROPFLG consistently outperforms the benchmark full OPF (FOPF) and the other two ROPF methods, achieving significant computational time savings while reliably finding optimal solutions.

## I. Introduction

Optimal Power Flow (OPF) is an essential tool in power system operations, analysis, and scheduling, including real-time economic dispatch. It involves optimizing an objective function, such as minimizing generation, while adhering to a set of physical and operational constraints, like generation limits and line thermal ratings. The scope of OPF applications has significantly expanded due to its ability to address various challenges within the power network, from grid management to resource allocation.

With the rapid growth of renewable energy integration, particularly from sources like wind and solar, the intermittent nature of these resources introduces new complexities to the system. Generation capacity can fluctuate dramatically throughout the day, making it crucial for OPF calculations to be both fast and highly accurate to ensure demand is met without compromising the stability or reliability of the grid.

To meet these evolving challenges, existing OPF models must be improved. This includes refining algorithms to handle the uncertainty and variability of renewable energy, enhancing computational efficiency for real-time application, and ensuring scalability as power systems grow more complex. Advances in OPF are vital to supporting the transition to cleaner energy while maintaining the reliable operation of the power grid.

Machine learning (ML) has been widely applied to address power flow problems [1]. Graph neural network (GNN) is an advanced neural network architecture tailored for graph-structured data, leveraging message passing between nodes to capture complex dependencies. The topology of the network is essential in GNNs, as it allows information to flow between adjacent nodes and edges through multiple layers, capturing both local and global relationships within the graph [2] [3]. GNNs have been implemented in various types of OPF problems [4] [5]. For instance, in [6], GNNs are used to predict $N$-1 contingencies, reducing computational time for solution discovery. Additionally, GNNs have been applied to network reconfigured OPF problems, where they optimize network topology to enhance solution efficiency [7].

Modifying the topology of a transmission network through techniques, such as line switching and bus splitting, has been previously employed as a corrective mechanism to mitigate congestion issues [8] [9]. In the context of transmission expansion planning, bus splitting has demonstrated its efficacy in enhancing optimal dispatch solutions [10]. The process involves physically altering the system topology by subdividing a single node into multiple nodes [11].

Network topologies of power systems are commonly represented as homogeneous graphs, where all nodes and edges are of a single type. Homogeneous graph restricts the GNN model from capturing distinct node features when multiple generators are associated with a single node. By subdividing virtual nodes from existing physical nodes, additional generator-specific features can be incorporated into the graph network. Since GNNs are highly sensitive to network topology, virtual node-splitting serves as an effective technique for enhancing network structure. Using virtual node-splitting, attributes for each individual generator such as generation costs, generation limits, and ramping rates can be utilized to extract critical information, thereby improving learning capacity and efficiency of the GNN model.

In our prior work, we developed an ML algorithm utilizing GNNs to improve the formulation of a reduced optimal power flow (ROPF) problem [12]. The GNN model was tailored to predict congested transmission lines under varying load profiles, enabling the identification of a critical subset of lines for active monitoring. By concentrating only on these critical lines, the approach reduces the number of line flow constraints within the OPF problem, effectively simplifying the problem and enhancing the efficiency of the ROPF formulation. The method not only accelerates OPF computations but also supports more scalable solutions, which are especially beneficial for systems with high variabilities in load profiles and generation capabilities, where real-time performance is essential.

In this paper, we present further advancements in the formulation of the ROPF problem, with key contributions and innovations summarized as follows:

- In contrast to the previous GNN model, which combined the capacities of all generators at each node as a single feature, this research employs the proposed virtual node-splitting strategy that enables GNN to explicitly model each generator individually. The approach substantially increases the number of node's features available for learning, thereby improving the accuracy of GNN predictions.
- An adaptive hierarchical GNN model is developed to predict generators that would operate at their maximum capacities, effectively reducing the number of generation variables and constraints and expediting the OPF solution process.
- Three ROPF methods—(i) ROPFL with reduced monitoring lines that are predicted to be congested, (ii) ROPFG with reduced variables and constraints related to the generators that are predicted to operate at their maximum power ratings, and (iii) ROPFLG with both reduced monitor lines and reduced generator-specific variables and constraints—were developed and benchmarked against the baseline full OPF (FOPF). Case studies demonstrate that the proposed ROPFLG consistently outperforms the benchmark FOPF as well as the other two ROPF methods. Leveraging the hierarchical GNN model, the proposed ROPFLG method achieves substantial reductions in computation time while consistently identifying the optimal solution.

## II. Background

### A. Reduced Optimal Power Flow

The OPF model is commonly used for real-time economic dispatch in power systems, optimizing generation dispatch to meet system loads while satisfying various constraints within the transmission network. Typically, the OPF objective function aims to minimize operating costs. In the OPF problem below, the total generation cost is minimized as expressed in (1), with decision variables representing generation $P_g$ and the corresponding cost $c_g$, while subject to constraints (2)–(5).

$$\textbf{Obj}: min \ \sum_{g \in G} c_g P_g \qquad g \in G \qquad (1)$$
$$P_g^{min} \le P_g \le P_g^{max} \qquad g \in G \qquad (2)$$
$$P_k = (\theta_{f(k)} - \theta_{t(k)})/x_k \qquad k \in K \qquad (3)$$
$$-RateA_k \le P_k \le RateA_k \qquad k \in K \qquad (4)$$
$$\sum_{g \in G(n)} P_g + \sum_{k \in K(n-)} P_k - \sum_{k \in K(n+)} P_k = d_n \qquad n \in N \qquad (5)$$

Constraint (2) establishes the minimum and maximum generation limits for each generator within the system. The power flow $P_k$ on each transmission line is calculated using (3). The line limit constraint in (4) restricts $P_k$ to be within the line's thermal rating limit, $RateA_k$. In constraint (3), $\theta_{f(k)}$ and $\theta_{t(k)}$ represent the phase angles at the sending and receiving ends of line $k$. Constraint (5) enforces nodal power balance, ensuring that the power generation meets the given nodal load $d_n$ at each bus.

In most OPF problems, the optimal solution is typically limited by a number of congested transmission lines. Independent System Operators (ISOs) often monitor a specific subset of lines that experience high loading or congestion. These subsets of lines are selected based on historical OPF solutions and load profiles. Similarly, maximum generation limits impose constraints on certain generators. Identifying generators that consistently operate at maximum capacity allows for a reduction in the total number of constraints.

Our previous work focused on dynamically selecting these critical lines in response to specific load profiles, utilizing machine learning, particularly GNN. In this study, we extend this approach by proposing a GNN model to predict which generators will reach maximum capacity under varying load conditions.

The hierarchical GNN model is designed to accurately forecast both congested lines and generators at their maximum generation for dynamic load profiles. By applying GNN predictions for line flows and generation levels, we can significantly reduce the number of constraints. The approach requires monitoring only a smaller subset of critical lines and dispatching a selected number of online generators within the OPF problem, resulting in a ROPF problem that alleviates real-time computational demands. Based on the predictions from the hierarchical GNN model, constraints for non-congested lines are ignored and parameters for maximum capacity generators are fixed during the formulation stage of ROPF problems. Ultimately, instead of applying constraints (2) and (4) across all lines and generators, we implement constraints (2a) and (4a) for a subset of congested lines $X$ and generators with undetermined generation levels $Y$.

$$P_y^{min} \le P_y \le P_y^{max} \qquad y \in Y \qquad (2a)$$
$$-RateA_x \le P_x \le RateA_x \qquad x \in X \qquad (4a)$$

By incorporating predictions from the GNN model, the ROPF solution identifies the optimal combination of generation outputs to satisfy the given load at minimal cost, while adhering to line limits and maintaining nodal power balance constraints. The GNN model can be updated in near-real-time with varying load profiles across different scenarios, enabling system operators to achieve optimal generation dispatch within the allocated computing time.

### B. Virtual Node-Splitting

Machine learning algorithms have been widely applied in power system optimization [13] [14]. However, these models often overlook the inherent topological information within electrical networks that could significantly improve their learning efficiency. GNN, a specialized class of ML models, address this by optimizing transformations across all graph components—nodes, edges, and global context—while maintaining graph symmetries, permutation invariance.

The unique capabilities of graph neural networks serve as powerful tools for advancing power system optimization by capturing and modeling the complex interdependencies within networked systems. Unlike other ML methods, GNNs leverage the structural properties of the network, enabling them to capture both local and global relationships across nodes and edges, which is essential for accurately predicting critical variables such as line flows and generation levels. The ability to integrate network topology into the learning process offers substantial improvements in prediction accuracy and computational efficiency.

In GNNs, network topology is incorporated through an adjacency matrix, an *nb* by *nb* matrix, where *nb* represents the number of buses. The adjacency matrix facilitates a more comprehensive representation of node and edge features. In most electrical networks, topology encodes essential, often implicit features that can be learned and extracted to enhance the model's performance.

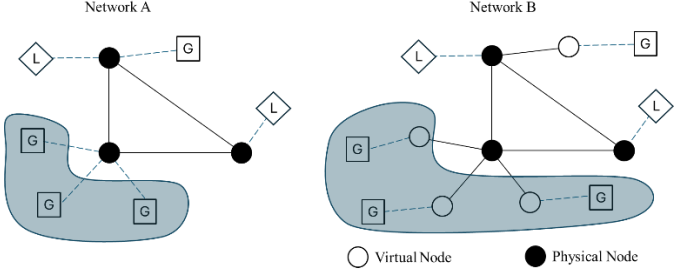

Fig. 1. A 3-bus system (Network A) vs. the expanded 3-bus system using virtual node-splitting (Network B).

As shown in Fig. 1, a simple 3-bus power system in Network A was augmented using virtual node-splitting and transformed into Network B. In our previous work with a homogeneous graph structure, the number of features per node or edge was restricted, preventing individual generator modeling at the node level. Generators at each bus were (e.g., the three generators in the shaded region) aggregated into a single equivalent generator, representing total generation at that node. Consequently, node-level generator-specific features, such as generation cost and ramp rate, were discarded. By using a virtual node-splitting, each generator can be represented as connected to a virtual node (white circle), while loads remain in the real nodes (black circle), allowing all generator-specific features to be used in GNN training.

GNNs leverage message passing, made possible by the adjacency matrix. Introducing virtual nodes increases the total number of nodes to *(nb + ng)*, where *ng* is the number of generators. The adjacency matrix is expanded into size of *(nb + ng)* by *(nb + ng)*. Although a larger GNN model with an expanded adjacency matrix may increase training time, it is expected to enhance learning performance. Once the GNN model is trained, predictions are made quickly. In GNNs, the more features provided per node, the more effectively the model can learn the network's global context, establishing relationships among various node and edge features.

By providing faster and more precise solutions, GNNs are particularly valuable for tackling the complex optimization challenges faced in modern power systems, such as those posed by high penetration of renewable energy sources and dynamic load conditions. As these networks evolve in scale and complexity, GNNs offer a complementary enhancement to traditional methods, using the inherent graph structure to gain deeper insights, streamline computation, and improve the robustness of optimization outcomes. Utilizing virtual nodes to expanded network topology for training GNN models holds promise for significantly improving operational efficiency and scalability in real-time applications within the power industry.

III. METHODOLOGIES

The GNN models and OPF algorithms were implemented in Python, utilizing the pyomo library and the gurobi solver for solving OPF problems [15]. The GNN models were developed with the spektral library [16], while figures and graphs were generated using matplotlib library. All computational tasks were performed on a system with a 4-core Intel i7-4790K CPU, a GTX 3090 GPU, and 32 GB of memory. To create a robust dataset for training and testing the GNN model, we conducted full OPF simulations on the IEEE 73-bus system under varied load profiles, generating 10,000 samples. For each sample, the load profile was adjusted within ±10% of the base load profile to ensure sufficient variability.

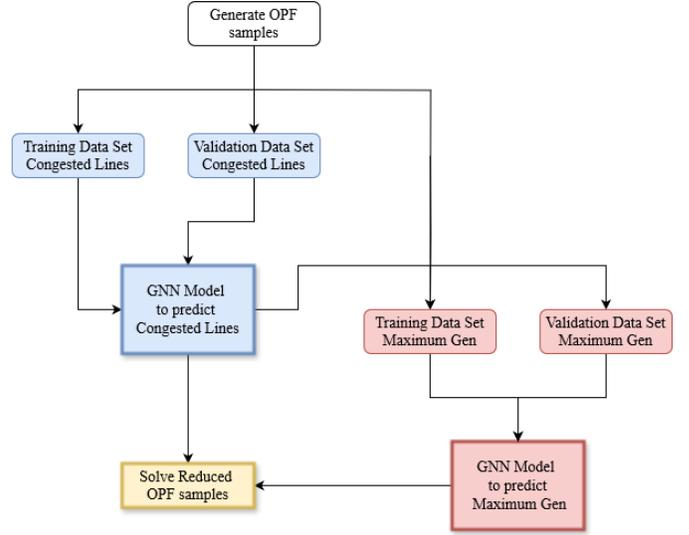

Fig. 2. Flow chart of proposed hierarchical GNN using virtual node-splitting.

A. OFFLINE – Training Stage:

The generated dataset, consisting of 10,000 samples, is partitioned into training and validation subsets. Labels for each transmission line are assigned based on a predetermined loading threshold, defined as a percentage of the line's thermal rating. For instance, if the thermal rating of a line is 200 MW and the loading threshold is set at 70%, the line is classified as congested when the power flow exceeds 140 MW. From Fig. 2, the first GNN model was trained to predict congested lines. Predictions for congested lines from the first GNN model will then be used as features to train the second GNN model. The second GNN model is then used to predict which generator will produce at maximum capacity. By knowing which lines will congest and which generators will produce at maximum capacity, the number of constraints in the FOPF problem can be significantly reduced resulting in a ROPF problem. The ROPF problem is then solved, and the solution is verified that all constraints have been met.

B. ONLINE – Operational Stage:

In the operational stage, the trained GNN models were deployed to process real-time test samples, predicting both congested lines and maximum-capacity generators. Four methods were implemented:

- **FOPF method**: the OPF problem is solved without reducing the number of constraints, monitoring all lines and dispatching all online generators, which serves as a baseline.
- **ROPFL method**: the GNN model predicts congested lines. Lines limit constraints for non-congested lines are removed, resulting in a ROPF problem.
- **ROPFG method**: the GNN model predicts maximum-capacity generators. Parameters for maximum-capacity generators are set, resulting in a different ROPF problem.
- **Proposed ROPFLG method**: the hierarchical approach uses the GNN model to first predict congested lines. These line predictions are then included as input features for the second GNN model, which predicts the generators that produce at their maximum capacity. Line limit constraints for non-congested lines are removed, and variables for predicted maximum-capacity generators are switched to known parameters that would reduce both variables and constraints, resulting in a further smaller ROPF problem.

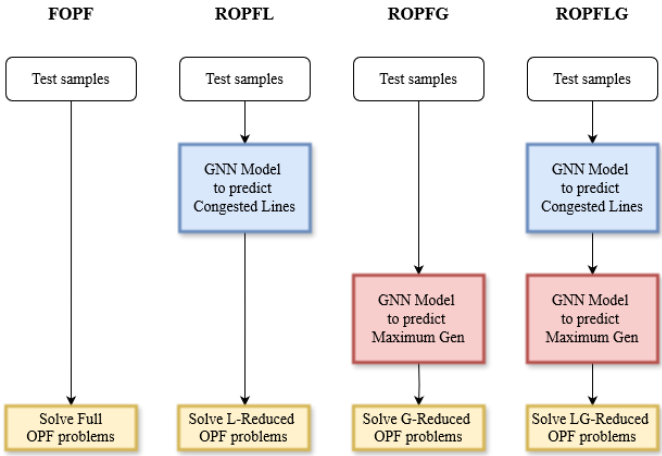

Fig. 3. Flow chart of three proposed ROPF and FOPF.

The flowchart in Fig. 3 summarizes the four proposed methods. Once the ROPF problem is formulated, it is solved, and its solution is verified for validity and optimality. By narrowing the focus to a smaller subset of critical constraints, the ROPF problem significantly decreases computational complexity, potentially enabling faster and more efficient solutions. The solving time of each ROPF method is then compared to that of the baseline FOPF method to assess performance improvements.

## IV. RESULTS AND ANALYSIS

This section evaluates the hierarchical GNN model's performance during training. As shown in Fig. 4,, the model demonstrates strong performance over 100 training epochs, with training accuracy closely tracking validation accuracy. Both figures confirm efficient training. The close alignment of training and validation loss curves in Fig. 5 suggests effective generalization to unseen data, minimizing overfitting risk. The smooth, consistent decrease in loss, eventually plateauing, indicates stable learning and convergence. Meanwhile, accuracy steadily rises during training, reaching approximately 96%. Overall, these results demonstrate the training process can effectively balance high accuracy with minimized loss.

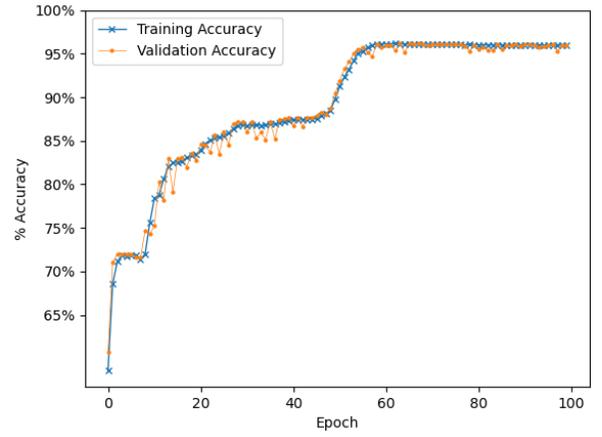

Fig. 4. Percent accuracy of prediction for training vs. validation for the GNN - Maximum Gen model.

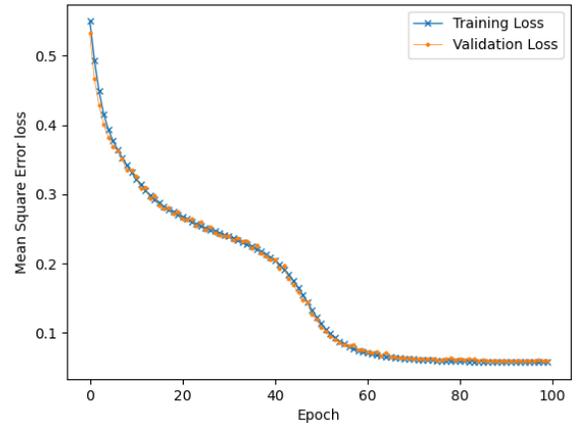

Fig. 5. MSE loss for training vs. validation for the for the GNN - Maximum Gen model

The trained GNN models were evaluated on a testing dataset of 1,000 samples, generating predictions for each sample regarding congested lines and maximum-capacity generators. These predictions present two potential error types:
- Type I error (false positive): Incorrectly classifying a non-congested line as congested or a generator as operating at maximum capacity when it is not.
- Type II error (false negative): Failing to identify an actual congested line or a generator operating at maximum capacity.

Previous work on ROPF indicate that solutions with a high Type II error rate (false negatives) are more likely to violate system constraints. There is a strong correlation between missing predictions for congested lines and maximum-capacity generator that can lead to invalid solutions for ROPF problem.

TABLE I. PERCENT ERROR FOR FALSE PREDICTIONS OF CONGESTED LINES AND MAXIMUM GENERATORS

|  | False Positive | False Negative | Total Error |
| --- | --- | --- | --- |
| Congested Lines | 1.07% | 0.12% | 1.19% |
| Maximum Generator | 5.35% | 0.92% | 6.27% |

Since predictions for congested lines serve as features for predicting maximum-capacity generators, a small error rate in congested line predictions may slightly increase the error rate for generator capacity predictions. As shown in TABLE I, the overall error rate for predicting maximum-capacity generators is slightly elevated at over 6%, but the false negative rate remains low, under 1%. False negative errors for predictions of congested line are excellent, with an error rate of approximately 0.1%. With these low false negative rates for both congested lines and maximum-capacity generators, the likelihood of generating invalid ROPF solutions from these predictions is expected to be minimal.

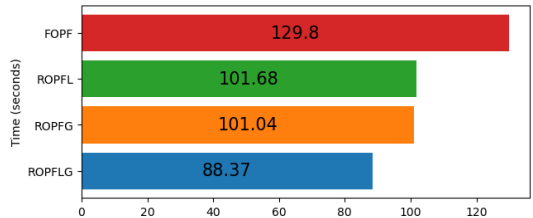

Fig. 6. Solving time using the four methods for 1,000 test samples.

In Fig. 6 and TABLE II, it compares the computation times for solving 1,000 OPF problems using the four methods. The FOPF method serves as the baseline, taking 129.80 seconds, which is set at 100%. The ROPFL method reduces the computation time to 101.68 seconds, achieving almost 22% of saving in solving time. Similarly, the ROPFG method takes 101.04 seconds, reaching a similar percentage of time saved. The most efficient method, ROPFLG, completes in 88.37 seconds, which is 32% faster than FOPF. The average total cost for 1000 samples is also shown in TABLE II. For the three ROPF methods, the mean total cost is almost identical to the benchmark FOPF model, with ROPFL and ROPFG slightly higher than ROPFLG; the difference is well below 0.1% which is negligible.

TABLE II.  MEAN TOTAL COST AND TIME SAVE IN (%) FOR SOLVING 1,000 TEST SAMPLES

|  | Mean Total Cost (%) | Time Saving (%) |
| --- | --- | --- |
| FOPF | 100% | 0 |
| ROPFL | 100.061% | 21.67% |
| ROPFG | 100.064% | 22.16% |
| **ROPFLG** | 100% | 31.92% |

The computational savings achieved with the hierarchical ROPFLG method, though substantial, are somewhat less than anticipated. It can be attributed to the limited number of constraints that can be removed before rendering the OPF problem unsolvable. Additionally, the computational benefits of ROPFLG are expected to be more pronounced in larger systems featuring multiple congested lines and several generators operating at maximum capacity. Overall, ROPFLG achieves the most significant reduction in computation time compared to other methods while maintaining the same optimal solution as the baseline FOPF.

## V. CONCLUSION

Among the four methods assessed in this study, the ROPFLG approach demonstrates the greatest computational time savings in finding solutions for OPF problems. All solutions obtained using the ROPFLG method are optimal, with no violations of operational constraints. Furthermore, the objective cost for each solution aligns closely with that of the baseline FOPF. Although the chosen ROPFLG method is more complex than FOPF—requiring the training of a hierarchical GNN model—the efficiency gains in solution speed are significant during the operational stage, especially for large-scale OPF problems.

The study also opens pathways for further exploration of GNN model variations. For instance, a transfer learning GNN model could be investigated, where the congested line GNN model is leveraged to predict maximum-capacity generators without additional retraining. Alternatively, a dual prediction GNN model could be developed to concurrently predict congested lines and maximum-capacity generators. Additionally, virtual node-splitting could be applied to address various OPF problem variants, such as *N*-1 contingency OPF or network-reconfigured OPF. Virtual node-splitting allows for the inclusion of more node-level features, thereby enhancing the model's capacity to learn and extract critical information, which ultimately improves prediction performance.

## VI. REFERENCES

[1] T. Pham and X. Li, "Neural Network-based Power Flow Model," in *IEEE Green Technologies Conference*, Houston, 2022.

[2] W. Liao, B. Bak-Jensen, J. R. Pillai, Y. Wang and Y. Wang, "A Review of Graph Neural Networks and Their Applications in Power Systems," *Journal of Modern Power Systems and Clean Energy,* vol. 10, no. 2, pp. 345 - 360, March 2022.

[3] J. Zhou, G. Cui, S. Hu, Z. Zhang, C. Yang, Z. Liu, L. Wang, C. Li and M. Sun, "Graph neural networks: A review of methods and applications," *AI Open 2021,* vol. 1, pp. 57-81, 2020.

[4] M. Tuo, X. Li and T. Zhao, "Graph Neural Network-based Power Flow Model," in *North American Power Symposium*, Asheville, 2023.

[5] A. V. Ramesh and X. Li, "Machine Learning Assisted Approach for Security-Constrained Unit Commitment," arXiv:2111.09824, Nov. 2021.

[6] T. Pham and X. Li, "N-1 Reduced Optimal Power Flow Using Augmented Hierarchical Graph Neural Network," arXiv preprint, 2024. [Online]. Available: https://arxiv.org/abs/2402.06226.

[7] T. Pham and X. Li, "Graph Neural Network-Accelerated Network-Reconfigured Optimal Power Flow," arXiv preprint, [Online]. Available: https://arxiv.org/abs/2410.17460.

[8] A. Hinneck, B. Morsy, D. Pozo and J. Bialek, "Optimal Power Flow with Substation Reconfiguration," in *2021 IEEE Madrid PowerTech*, Madrid, 2021.

[9] M. Heidarifar, M. Doostizadeh and H. Ghasemi, "Optimal transmission reconfiguration through line switching and bus splitting," in *2014 IEEE PES General Meeting*, National Harbor, 2014.

[10] P. J. N. Gealone and A. E. D. Tio, "Optimized Transmission Expansion Planning with Bus-Splitting in Grids with High VRE Penetration," in *SIEDS*, Charlottesville, 2024.

[11] M. D. Librandi, D. Stenzel, R. Witzmann, H. Hitzeroth, N. Kaes and S. Bopp, "Towards a Set of Bus-Splitting Topologies for Congestion Management," in *NEIS*, Hamburg, 2023.

[12] T. Pham and X. Li, "Reduced Optimal Power Flow Using Graph Neural Network," in *2022 North American Power Symposium*, Salt Lake City, 2022.

[13] M. Tuo and X. Li, "Long-term Recurrent Convolutional Networks-based Inertia Estimation using Ambient Measurements," in *IEEE IAS Annual Meeting*, Detroit, MI, USA, 2022.

[14] A. V. Ramesh and X. Li, "Spatio-Temporal Deep Learning-Assisted Reduced Security-Constrained Unit Commitment," *IEEE Transactions on Power Systems,* 2023.

[15] W. E. Hart, J.-P. Watson and D. L. Woodruff, "Pyomo: modeling and solving mathematical programs in Python," *Mathematical Programming Computation,* vol. 3, pp. 219-260, 2011.

[16] J. B. Maguire, D. Grattarola, V. Mulligan, E. Klyshko and H. Melo, "XENet: Using a new graph convolution to accelerate the timeline for protein design on quantum computers," *PLoS Computational Biology,* no. September, 2021.